# The Heats of Dilution. Calorimetry and Van't-Hoff.

I. A. Stepanov

Latvian University, Rainis bulv. 19, Riga, LV-1586, Latvia

e-mail istepanov@email.com

## Abstract

Earlier it has been found that there is a big difference between heats of dilution measured by calorimetry and by the Van't-Hoff equation. In the present paper a reason for that is proposed. Experimental data for dilution of benzene and n-hexane in water were used.

## 1. Introduction

There is a big difference between heats of dilution measured by calorimetry and by the Van't-Hoff equation. Let's cite [1]: "Historically, the Van't-Hoff isochore has frequently been used as a means of deriving enthalpy change values for reactions of interest from experimental data gathered at a number of temperatures. ... Detailed review of the data in articles for reactive systems where the enthalpy change has been determined both by calorimetric measurement and other temperature-change techniques show that there is often a sizeable discrepancy between the two values. ... It is possible to show that improvements in instrumentation and computing power have not overcome earlier difficulties in the estimation of enthalpy values from data taken at several temperatures. Perusal of literature over the last three decades shows that many authors have been aware that obtaining enthalpy values by these techniques is inherently unsatisfactory". In private communication to the author the author of [1] wrote: "There are a number of comprehensive collections of enthalpy data, particularly those authored by R. M. Izatt and J. J. Christensen, which contain all the thermodynamic data for many systems. Further, many of these systems contain data collected by calorimetric determination and data from 1/T estimation methods in the same table. As I examined those it seemed that there was always a consistent difference between the two methods".

In the present paper an attempt is made to explain this papadox.

## 2. Theory

For chemical processes the law of conservation of energy is written in the following form:

$$dU = dQ - PdV + \sum_i \mu_i dN_i \qquad (1)$$

where $dQ$ is the heat of reaction, $dU$ is the change in the internal energy, $\mu_i$ are chemical potentials and $dN_i$ are the changes in the number of moles.

In [2-6] it has been shown that the energy balance in the form of (1) for the biggest part of the chemical reactions is not correct. In the biggest part of the chemical reactions the law of conservation of energy must have the following form:

$$dU = dQ + PdV + \sum_i \mu_i dN_i \qquad (2)$$

The Van't-Hoff equation is the following one:

$$\frac{d}{dT}\ln K = \Delta H^0/RT^2 \qquad (3)$$

where K is the reaction equilibrium constant and $\Delta H^0$ is the enthalpy. According to thermodynamics, the Van't-Hoff equation must give the same results as calorimetry because it is derived from the 1st and the 2nd law of thermodynamics without simplifications. However, there is a paradox: the heat of chemical reactions, that of dilution of liquids and that of other chemical processes measured by calorimetry and by the Van't-Hoff equation differ significantly [1-6]. The difference is far beyond the error limits. The reason is that in the derivation of the Van't-Hoff equation it is necessary to take into account the law of conservation in the form of (2), not of (1) [2-6].

If to derive the Van't-Hoff equation using (2) the result will be the following one:

$$\frac{d}{dT}\ln K = \Delta H^{0*}/RT^2 \qquad (4)$$

where $\Delta H^{0*} = \Delta Q^0 + P\Delta V^0$.

## 3. Experimental Check and Discussion

In [7] heats of dilution of benzene and n-hexane in water were given (Table 1). One sees that $\Delta H$ is a few times bigger than the experimental value $\Delta Q$. It is impossible to explain that by non-ideality of solution. In [7-9] they use $K_c$ and $K_a$ in (3) (the constants depending on the concentrations and on the activities, respectively). It confirms the result [2-6] that in chemical processes $\Delta Q$ is not equal to $\Delta H$.

From Table 1 it is possible to find the change in the volume $\Delta V$ for dilution. For benzene $\Delta V \approx 10^{-2}$ and $2 \cdot 10^{-2}$ m$^3$/mol, for n-hexane $\Delta V \approx 2 \cdot 10^{-2}$ m$^3$/mol. $\Delta V = V_2 - V_1 \approx V_2$ because for both substances $V_1 \ll \Delta V$. Pay attention that for both substances $V_2$ is close to the volume of the ideal gas ($2,5 \cdot 10^{-2}$ m$^3$/mol).

Let's analyze a phase transition which is not a chemical reaction [10]

$$Zn(liq)=Zn(gas) \qquad 700<T<1000 \text{ K.} \qquad (5)$$

$$\ln P = -118366/RT + 12,049 \qquad (6)$$

From (6) $\Delta H^0=118,366$ kJ/mol. From [11] for T=900 $\Delta Q=118,31$ kJ/mol, for T=800 $\Delta Q=119,37$ kJ/mol. One sees that in this case the usual Van't-Hoff equation is valid.

The Van't-Hoff equation (3) was often used for determination of the heat of reaction. It gives correct results for reactions with $\Delta V \to 0$. Therefore, it was assumed that for chemical reactions

$$\Delta Q = \Delta H = \Delta U + P\Delta V \qquad (7)$$

However from (1) and (7) it follows

$$\Delta Q = \Delta Q + \sum \mu_i \Delta N_i \qquad (8)$$

This conclusion is an absurd: one can not neglect the last term in (8). Therefore, the Van't-Hoff equation must not give correct results neither for $\Delta V \to 0$ nor for $\Delta V \neq 0$. For reactions with $\Delta V \neq 0$ this equation gives wrong results [1-6]. The present theory explains this paradox: in chemical processes (4) must be used.

# References

1. P. R. Brown, in *Proc. 12th IUPAC Conf. Chem. Thermodynamics*, Snowbird, Utah, August, 1992, 238-239.

2. I. A. Stepanov, *DEP VINITI*, No 37-B96, (1996). Available from VINITI, Moscow.

3. I. A. Stepanov, *DEP VINITI*, No 3387-B98. (1998). Available from VINITI, Moscow.

4. I. A. Stepanov, *7$^{th}$ European Symposium on Thermal Analysis and Calorimetry.* Aug. 30 - Sept. 4. Balatonfuered, Hungary. 1998. Book of Abstracts. P. 402-403.

5. I. A. Stepanov, The Law of Conservation of Energy in Chemical Reactions.- *http://ArXiv.org/abs/physics/0010052.*

6. I. A. Stepanov, The Heats of Reactions. Calorimetry and Van't-Hoff. 1. - *http://ArXiv.org/abs/physics/0010054.*

7. D. S. Reid, M. A. Quickenden, F. Franks, *Nature*, 224 (1969) 1293.

8. C. V. Krishnan, H. L. Friedman, *J. Phys. Chem.*, 73 (1969) 1572.

9. R. L. Bohon, W. F. Claussen, *J. Amer. Chem. Soc.*, 73 (1951) 1571.

10. R. Hultgren, P. Desai, D. Hawkins, M. Gleiser, K. Kelly, *Selected Values of the Thermodynamic Properties of the Elements*, (American Society for Metals, Metals Park, OH) 1973.

11. I. Barin, *Thermochemical Data of Pure Substances* (VCH, Weinheim, Germany) 1989.

**Table 1**

Heats of dilution of benzene and n-hexane in water [7]

|  | $\Delta Q$/J/mol, Calorimetry | $\Delta H$/J/mol, Van't-Hoff |
|---|---|---|
| **Benzene** | 800<br>460 [8] | 1820<br>2430 [9] |
| **n-hexane** | 470 | 2500 |